# Locating transition path region in the free energy landscape of protein folding

Debajyoti De, Anurag Singh and Amar Nath Gupta*

Biophysics and Soft Matter Laboratory, Department of Physics, IIT Kharagpur, Kharagpur-721302, India

*Corresponding author's email id: ang@phy.iitkgp.ernet.in

## Abstract

Protein folding processes are generally described statistically with the help of multidimensional free energy landscape, typically reduced to a 1-D free energy profile along good reaction co-ordinate. There are many physical parameters which are responsible for protein molecule to hop between the native and unfolded states. The transition path region across the barrier is the region corresponding to the minimum fluctuation. The extent to which this transition region can extend beyond the obvious ½ $k_B T$ region has been a question of interest for a long time. We propose a new method to locate this transition path region and to study its dependence on the asymmetry of the transition state for a given free energy landscape. We have performed Brownian dynamics simulations with Gaussian white noise and Monte-Carlo simulation by sampling ten thousand successful transitions across the barrier for three different energy landscape having fixed barrier height with asymmetry in their curvatures. It was found that the transition path region increases with the increase in the asymmetry of the energy landscape in the transition region. The rate limiting parameters - diffusion constant over the diffusive barrier, rate constant at particular force and transition path time for different potentials were estimated directly from the landscape profile using Kramers' theory for diffusive barrier crossing. It was also found that the average diffusion in the native state increases with the increase in the asymmetry of the transition state towards the non-native state.

## Introduction

Kinetics of protein folding at single-molecule level is generally studied through Kramer's theory and transition rate theory using free energy landscape [1]. The free energy landscape of protein contain several important information ranging from folding kinetics, location of transition states and existence of intermediates in the pathway of folding. The shape of the energy profile is crucial for estimation of rate limiting factor, which is diffusion across the barrier in the course of folding.

The free energy landscape also contains the information of the structure of the protein. The symmetry (asymmetry) in the free energy landscape depicts that the behaviour of the protein when transiting from native to unfolded and unfolded to native state is same (different). This indicates towards the symmetry (asymmetry) in the protein at the structural level. This structural asymmetry compels the protein to behave differently, even in the same environmental conditions, in comparison to the protein corresponding to the symmetric landscape. Previously, many studies have been done to get the kinetic properties of the protein in a given environmental condition [2, 3]. In such studies, the kinetic parameters depend on the environmental conditions in which they were estimated but, best to our knowledge, no attempt of comparison between different proteins in the same environmental conditions has been done till now.

Another important parameters related to protein folding are transition path region and transition path time. The transition paths are the small fraction of an equilibrium trajectory for successful transition between the states. These trajectories pass through the transition state and provide very useful information about the dynamics of biomolecular folding, how non-native structures are formed in the way of folding. The molecule, after being in these transition states, can move to any of the folded or unfolded state. The crossing of this region means the molecule has successfully transited from one state to the other state. The general understanding about the transition region in term of energy is ½ $k_B T$ which corresponds to the minimum thermal fluctuations present in the system. The duration of crossing this region is very short compared to the rate of folding. The direct measurement of the transition path time is extremely difficult because of the time resolution of the apparatus of single-molecule set up [AFM, Optical tweezers and Fluorescence spectroscopy]. Experimentally, the protein folding has been studied using Single molecule fluorescence spectroscopy (SMFS), single molecule FRET experiments, nuclear magnetic resonance (NMR) studies and hydrogen mass spectroscopy [4-7]. Also, the reconstruction of free energy landscapes of single-molecule protein by SMFS equilibrium as well as non-equilibrium trajectories has also been explored [8, 9]. These studies help us in understanding the behaviour of the protein in the experimental environment. However, no study has been done which depicts the variation of transition path region and transition path time for different kinds of free energy landscape.

The transition path time ($\tau_{tp}$) is insensitive to the barrier height of different proteins (two-state folding proteins), but sensitive to temperature and viscosity of the supporting medium [10]. It was shown that the variation in $\tau_{tp}$ comes from the variation in diffusion constant and not from the curvature of the well belonging to native state of protein in the free energy profile. The diffusion of protein molecule in the transition region is weakly dependent on the solvent viscosity, but intra-molecular interaction may be another possibility for the diffusive behaviour in this region.

The transition states are unstable intermediates through which the molecule passes to gain different conformations. The region of transition path across peak of the barrier is debatable. How this region can be located and their dependencies on asymmetry of the barrier is challenging. In this work we have chosen three different free energy profiles having only one transition state. We kept the barrier height same, but shifted the barrier position toward the non-native state to locate the transition path region and their dependencies on the asymmetry of the free energy landscape at constant force.

The 1-D free energy landscape can be manipulated by force keeping other parameters unchanged. It is very nice control parameter. The structured part of the protein molecule can be stretch out by the force and convert folding into motion. The energy landscape can be alter, tilted at any given force and study the property of that particular state. The structure of protein can be changed without changing the buffer conditions like temperature, salt, pH, denaturant. At a particular constant force ($F_{1/2}$), a single-molecule protein can hop between the folded and the unfolded state. While making a transition, the protein diffuses through various possible conformations along the reaction coordinate. The cause of this diffusion is the thermal fluctuations present in the system. This thermal noise essentially is the driving force for the Brownian dynamics of the protein folding. In past, this random process has been replicated by doing Brownian dynamics simulations in the presence of white Gaussian noise [11-13].

We have used two different methods to locate the region of interest. First, we have performed Brownian dynamic computer simulation with white noise as fluctuation and the second; we have used Monte-Carlo simulation by sampling 10 thousand successful transitions across the transition state. It was found that the $\tau_{tp}$ is very sensitive to the asymmetry of the transition state and it is wider than ½ $k_BT$ region depending on the asymmetry of the barrier. We have estimated rate limiting parameter diffusion constant over the diffusive barrier, rate constant at particular force and transition path time for different potentials directly from the landscape profile using Kramer's theory for diffusive barrier crossing. It was also found that the average diffusion in the native state increases with the increase in the asymmetry of the transition state towards the non-native state.

**Theory and Methods**

The diffusion constant of single-molecule is an important dynamical property which is responsible for change in configuration along the 1-D free energy profile against the good reaction co-ordinate [14]. The kinetic properties of protein, for example folding rate of single-molecule protein, entirely depend on the nature of the free energy landscape of the protein. The time evolution of a single-molecule along the free energy landscape in 1-D can be described by balancing the forces as [15],

$$m\ddot{x}(t) = -\gamma \dot{x}(t) - \frac{\partial U(x(t))}{\partial x} + \sqrt{\gamma k_B T}\zeta(t) \tag{1}$$

The Langevin's equation was obtained in the limit of low Reynolds number by dropping the inertial term $m\ddot{x}(t)$ from the above equation and can be written as

$$\dot{x}(t) = -\frac{1}{\gamma}\frac{\partial U(x(t))}{\partial x} + \zeta(t)\sqrt{\frac{k_B T}{\gamma}} \tag{2}$$

The time evolution of the molecule was obtained by integrating the above equation with respect to *t*,

$$x_2(t) = x_1(t) - \Delta t \left[\frac{1}{\gamma}\frac{\partial U(x(t))}{\partial x} - \zeta(t)\sqrt{\frac{k_B T}{\gamma}}\right] \tag{3}$$

where, $x(t)$ denotes the time-dependent position of the molecule, $\gamma$ is the frictional coefficient, $k_B$ is the Boltzmann constant, $T$ is the temperature, $U(x)$ is the potential function, and $\zeta(t)$ is white Gaussian noise. The thermal fluctuations at temperature $T$ were modelled by $\zeta(t)$ with zero mean and standard deviation of one, which obeys the Fluctuation-Dissipation relation.

The transition rate and path time (the time required to cross the barrier) has been estimated from the Kramer's theory by using free energy landscape parameters like the curvature of the wells, barrier height and their positions from the wells with the following expressions [16],

$$k_{u,0} = k_u \, exp(\Delta G^{\ddagger}/k_B T) \tag{4}$$

$$\tau_{tp} = \frac{\ln(2e^{Y}\Delta G^{\ddagger}/k_{B}T)}{2\pi k_{u,0}\sqrt{\varkappa_{b}/\varkappa_{w}}} \quad \text{Where,} \quad \Delta G^{\ddagger} > 2k_{B}T \tag{5}$$

Where, $k_u$ is the unfolding transition rate, $k_{u,0}$ is the unfolding transition rate at zero force, $\Delta G^{\ddagger}$ is the barrier height, $Y$ is Euler's constant, $\varkappa_w$ and $\varkappa_b$ are the stiffness (average curvature) of the well and barrier respectively.

The effective diffusion coefficient of the molecule can be expressed as,

$$D(t) = \lim_{t \to \infty} \frac{1}{2t}\langle(x(t) - x(0))^2\rangle \tag{6}$$

Where $x(0)$ is the initial position and $x(t)$ is the position of the molecule at time $t$. The Brownian dynamics simulation were done on three different potential function having equal barrier height of $10\,k_BT$, same well positions but different position of transition state, see Fig. 1. The parameter $\gamma$ was taken to be $1.0 \times 10^{-11}\,kg/s$ which include the viscosity of the medium as $1\,cp$ at room temperature. The calculated size of the protein molecule using $\gamma$ is $1.16\,nm$. The time interval of $\Delta t = 0.5\,ms$ was taken which is very close to the sample rate of most of the equilibrium experiments [7].

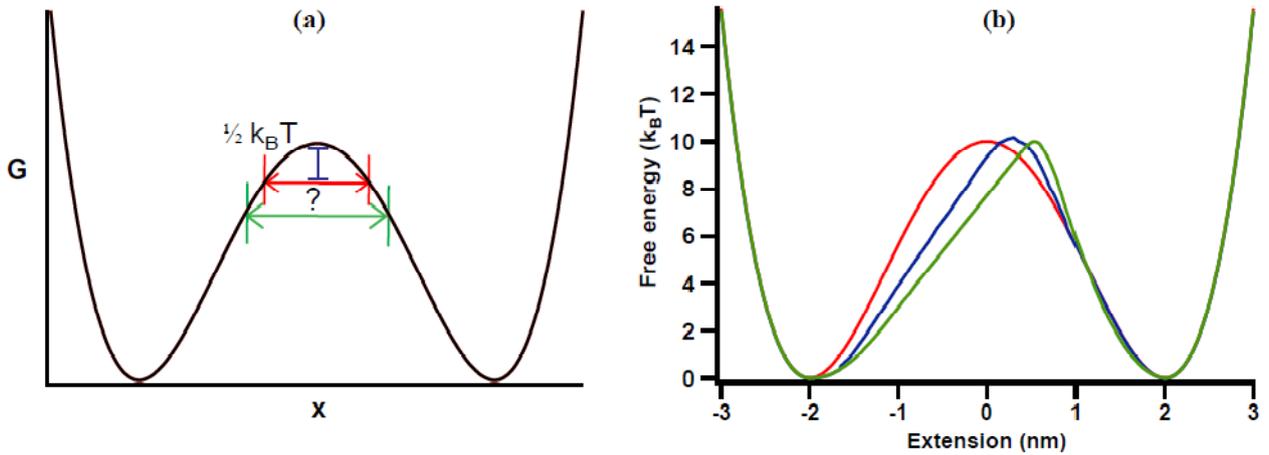

Figure 1: (a) Schematic representation of the transition path (green) across the barrier and how to locate this region compared to usual ½ $k_BT$ region (red). (b) Free-energy profile for the symmetric (red curve) and two asymmetric double well potentials (blue and green curve respectively). The well positioning and the barrier height of all the three potentials are same. Only the transition state of the asymmetric potentials are shifted with respect to the symmetric by 0.30 *nm* (blue curve) and 0.55 *nm* (green curve) respectively.

The considered potential functions belong to three different proteins having same size. By applying the constraints of same well positioning and same barrier height, we have assured that the possible cause of differences in the kinetic properties of these proteins is due to the differences in their behaviour when they are transiting between native and unfolded states. The symmetric potential for first protein indicates that the behaviour in presence of the Gaussian noise is same for both folding and unfolding transitions. Asymmetry introduced in the second and third potential indicates that the behaviour of the protein is different while it is making transitions between the native to unfolded state and unfolded to native state. The shift in transition state 0.30 nm and 0.55 nm from symmetric potential towards the non-native state appears by keeping the curvature of the non-native state unchanged. These shifts can be taken arbitrary.

## Results

In order to obtain the transition path region, we have divided the energy profile into two parts by taking the barrier peak as reference point. The first part corresponds to the native and second part corresponds to the non-native state of the protein. The transition path region is a small part of the energy profile across the transition state where protein spends very small amount of time. When protein molecule crosses this region and changes its state (native to unfolded or unfolded to native), we assign the transition as a successful transition. We collect all the pre and post transition points in the two regions and then we did statistical analysis to locate the transition path region.

*Monte-Carlo simulation:*

In the Monte-Carlo simulation, we have sampled trajectories for both unfolding and folding transitions. For unfolding (folding) transitions, we have chosen starting extension with x = -2 *nm* (x = 2 *nm*), which corresponds to native (unfolded) state of the protein, along the reaction coordinate and allowed the protein to diffuse until it makes a successful transition. We have only considered the extensions data points between the x=-2 *nm* and x=2 *nm*. This is mainly because we are interested in the conformations which are between native and unfolded state of the protein, as these are the conformations which are more physically relevant. We have then collected the pre-transition and post-transition point of this successful transition. We again repeat the process ten thousand times of such trajectories. The weighted average of pre-transition points corresponding to the unfolding and folding transitions gave us the required transition region for the given potential profile.

*Brownian dynamics simulation:*

The single-molecule constant force data from Brownian dynamics simulation were segregated into two parts: the trajectories which are starting in the range [-2.1, -1.9] and [1.9, 2.1] for unfolding and folding transitions and finally resulting in successful transitions. The pre-transition and post-transition points of these trajectories were collected. In this case also, the transition path region was estimated by the weighted average of pre-transition points corresponding to the unfolding and folding trajectories. The choice of the starting regions of these trajectories was done by fitting the wells corresponding to native and unfolded state of the protein with the harmonic functions. The points from where the harmonic function start leaving the potential profile gives us starting regions for these trajectories. The transition region does not depend much on the choice of this starting region for large number of trajectories in the ensemble (in case of Monte-Carlo) and for large time data (in case of Brownian dynamics), the transition region converges to a fixed value (see S1, supplementary text).

The distribution of the pre-transition points in the native and unfolded part of all three potentials is shown in Fig. 2. In case of symmetric potential, we have observed that the distributions of the pre-transition points in the native and unfolded parts of the potential are same irrespective of the methods (Monte-Carlo or Brownian dynamics) used (see Fig. 2 for symmetric).

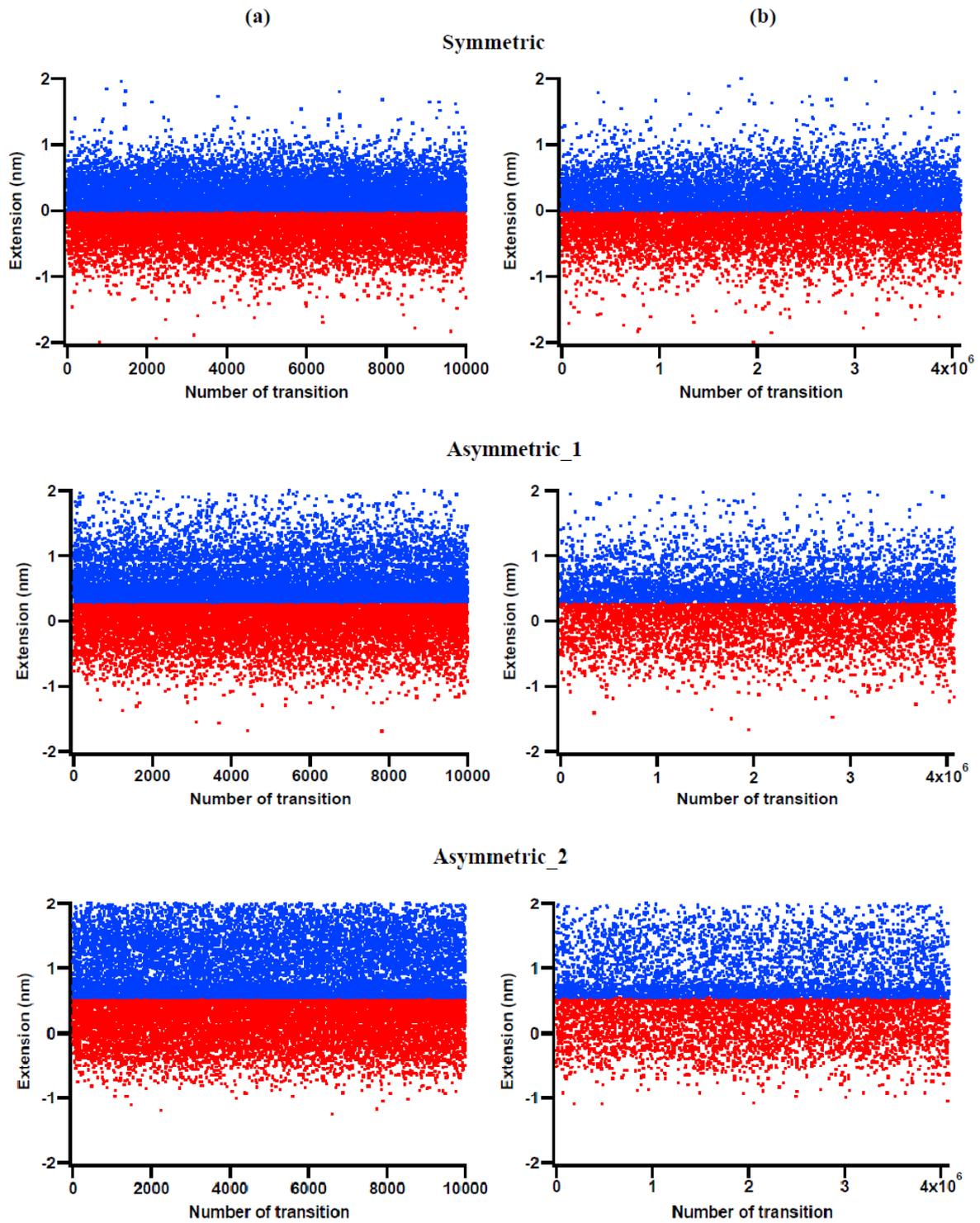

Figure 2: The panel (a) and (b) shows the distribution of pre-transition points in the native and unfolded state of the potential obtained using Monte-Carlo and Brownian dynamics respectively for all the three potentials. The red dots represent the pre-transition points corresponding to the unfolding transitions and the blue dots represent the pre-transition points corresponding to the folding transitions.

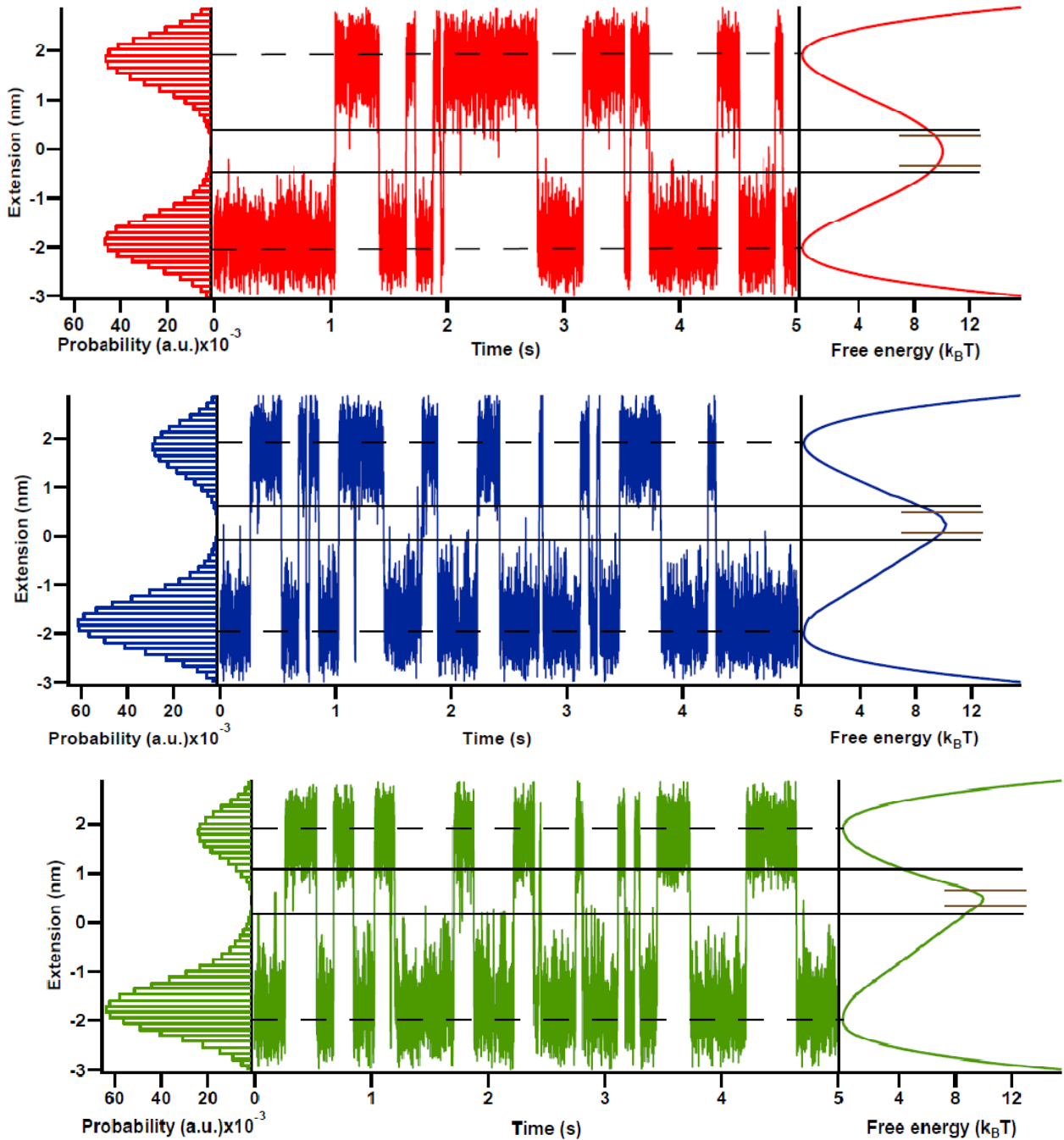

Figure 3: The results of the Brownian dynamics simulation for symmetric (red), asymmetric_1 (blue) and asymmetric_2 (green) double well potentials. The response of the molecule at a constant force ($F_{1/2}$) shows the abrupt jumps, as the molecule unfolds and refolds in the two-state potentials. Here we also plotted the corresponding probability distribution of the molecule. The dashed black line corresponds to native and non-native state of the protein molecule. The two black solid line locating the transition path region. The equivalent region is also shown on the potential. The brown line corresponds to the ½ $k_B T$ region.

This explains the symmetric nature of the transition region obtained in case of symmetric potential. After analyzing the distribution of pre and post transition points of asymmetric potentials we have found that the distributions of pre-transition points are biased towards the unfolded part of the potential (see Fig. 2 for asymmetric_1 and asymmetric_2). This explains the biasness in the transition region obtained in case of the asymmetric double well potentials. This biasness in the distribution of the pre-transition points may be because of the change in the steepness in the potential caused by the introduced asymmetry in the potentials. Also, these distributed points followed Gaussian distribution (in case of symmetric potential) and exponential distribution (in case of asymmetric potentials) having distribution parameters depending on the steepness in the potential (see S2, supplementary text).

As the asymmetry in the potential increases, the well belongs to native structure became wider and the well belongs to unfolded structure became narrower. Now the characteristic time spent in the wider well is large than the symmetric case. In the folded region protein has to cover a large distance to cross the transition region for a successful transition. This fact is supported by the distribution in the native and the unfolded state for all three potentials (see Fig. 2).

The reconstruction of free energy landscapes from Brownian dynamics data at constant force matches the potentials which were considered in the beginning. The native and non-native state was exactly located by fitting the extension histogram with Gaussian function for symmetric potential and skew Gaussian function for asymmetric potential which matches with the peak values of the both kinds of double well potentials. The solid black line corresponds to the transition path region which is shown on the three different potentials (see Fig. 3) from different simulation techniques.

The lifetime of the folded/native state was estimated by segregating the data between two successful transitions from native to non-native state. The distribution of lifetime of the folded states of protein at a particular constant force is shown in Fig. 4. It is obvious from the plot that the lifetime of the unfolded states increases as the number of counts decreases which means the rate of transition increases. The histogram plot of counts of transition v/s lifetime was fitted with a single exponential decay function to estimate unfolding rate, $k_u$ and which helps to determine the unfolding rate at zero force, $k_{0,u}$ (see Fig. 4).

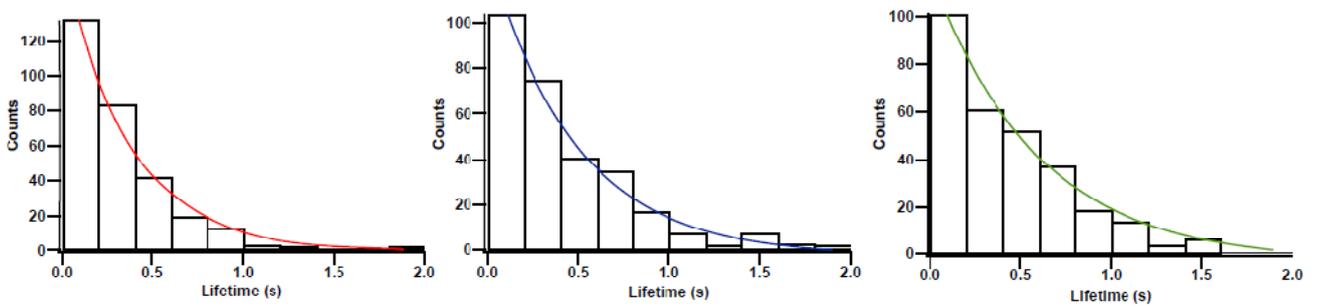

Figure 4: Count v/s lifetime plot is shown for native state of protein molecule having symmetric and asymmetric free energy landscapes. The histogram plot was fitted with the single exponential curve. The lifetime of native state follows the exponential distribution.

The barrier height ($\Delta G^{\ddagger}$) for the symmetric and asymmetric double-well potentials were fixed at $10\ k_B T$. The unfolding rates, $k_u$ for the three different protein molecule, which follows the symmetric and asymmetric potentials, are $2.74 \pm 0.18\ /s$, $2.04 \pm 0.24\ /s$ and $1.70 \pm 0.23\ /s$ respectively. Here the sources of error came from the fitting itself. The estimated unfolding rate at zero force $k_{0,u}$ was calculated using Eq. (4). The values of $k_{0,u}$ are $(6.04 \pm 0.41) \times 10^4\ /s$, $(4.5 \pm 0.53) \times 10^4\ /s$ and $(3.75 \pm 0.51) \times 10^4\ /s$) for symmetric and asymmetric potentials respectively. Another very important parameter of protein folding is transition path time ($\tau_{tp}$) which determines the speed of folding. The $\tau_{tp}$ is the time protein takes to cross the transition path region and this was estimated by using the parameters; folding rate at zero force ($k_{0,u}$), average curvatures of the native and transition state ($\varkappa_w$ and $\varkappa_b$), and barrier height ($\Delta G^{\ddagger}$), mentioned in Eq. (5). The estimated $\tau_p$ was obtained using Kramer's theory for both the symmetric and asymmetric double well potentials are $9.8 \pm 0.7\ \mu s$, $12.57 \pm 1.48\ \mu s$ and $14.25 \pm 1.94\ \mu s$ respectively. The diffusion constant which is an important parameter in protein folding for a given landscape was estimated from the Eq. (6) using the Brownian dynamics data. The values of the average apparent diffusion coefficients obtained are $309.3 \pm 1.29\ nm^2/s$, $394.78 \pm 1.45\ nm^2/s$ and $506.08 \pm 2.15\ nm^2/s$ for the three potentials respectively. The error in diffusion constant was estimated from different trajectories from performing Monte-Carlo simulation. It was observed that the value of the diffusion coefficient increases as the asymmetry in the potential increases. This is mainly because of the increased conformational freedom of the protein in the native state based on the level of asymmetry introduced in the potential. The diffusion coefficients obtained from Kramer's theory were in the order of $10^{-13}\ m^2/s$, which is two orders higher than average apparent diffusion obtained using the Brownian dynamics data. Previously, Fernandez et. al. [18] has measured the reconfiguration time for the unfolded protein, which gave the apparent diffusion coefficient in the order of $10^{-15}\ m^2/s$. Also, extremely slow intra-molecular diffusion had also been reported for unfolded proteins [19]. All the obtained values of the symmetric and the two asymmetric potentials are tabulated in table 1.

Table 1. The key parameters related to three different proteins (potentials) are presented here.

| Potentials/Parameters | Symmetric | Asymmetric_1 | Asymmetric_2 |
|---|---|---|---|
| Transition path region (in $k_BT$) | [0.58 , 0.58] | [1.26 , 1.76] | [2.07 , 3.61] |
| % increase wrt ½ $k_BT$ region using MC simulation | [16 , 16] | [152 , 252] | [314 , 622] |
| % increase wrt ½ $k_BT$ region using BD simulation | [16 , 16] | [160 , 146] | [308 , 850] |
| ½ $k_BT$ region (in $nm$) | [-0.32 , 0.32]=0.64 | [0.07 , 0.49]=0.41 | [0.39 , 0.66]=0.27 |
| TP region (in $nm$) using MC simulation | [-0.34 , 0.34]=0.68 | [-0.09 , 0.67]=0.76 | [0.12 , 1.07]=0.95 |
| TP region (in $nm$) using BD simulation | [-0.34 , 0.34]=0.68 | [-0.09 , 0.61]=0.70 | [0.12 , 0.95]=0.83 |
| % of data points in ½ $k_BT$ region | 0.96 | 0.57 | 0.39 |
| % of data points in TP region by MC simulations | 1.06 | 1.21 | 1.67 |
| % of data points in TP region by BD simulations | 1.06 | 1.67 | 2.26 |
| % of time in TP region by MC simulations | 1.05 | 1.23 | 2.99 |
| % of time in TP region by BD simulations | 1.05 | 1.08 | 2.13 |
| Unfolding rate (/s) at $F_{1/2}$ | 2.74 | 2.04 | 1.7 |
| Unfolding rate at zero force ($10^4$ /s) | 6.04 | 4.5 | 3.75 |
| Transition path time ($\mu s$) | 9.8 | 12.57 | 14.25 |
| Diffusion constant ($n\ m^2/s$) | 309.3 | 394.78 | 506.08 |
| TP region (in $nm$) using fitting method | [-0.45, 0.45]=0.9 | [0.06, 0.68]=0.62 | [0.09, 0.75]=0.66 |
| TP region (in $nm$) by rough estimation | [-0.48, 0.48]=0.96 | [-0.11, 0.77]=0.88 | [0.02, 0.93]=0.91 |

## Discussion

In past, two different methods were used for locating the possible transition path region. In the first approach, the part of the potential profile, which contained the possible transition region, was fitted with Gaussian or skew-Gaussian curve depending on the symmetric or asymmetric nature of the potential profile [20] and the transition region was given by the region from where the fitting curve started leaving the potential profile. In the second approach, a rough estimation of the transition path region was done by observing the transitions of the molecule in the extension v/s time data obtained from Brownian dynamics simulation [21]. The above two methods were also used in this work (see S3, supplementary text) and compared with the new methods to locate the transition path region (see table 1). The ambiguity in the fitting method arises from the selection of the point from where the fitting curve start leaving the potential curve. The selection of this point may differ based on the standard deviation of the Gaussian or skew-Gaussian fitting curves. Therefore the result is highly dependable on the standard deviation of the fitted curve. The problem with the second method is that the transition region is chosen based on the extension v/s time plot of the Brownian dynamics data. In such process, the choice of the region is customary in nature and only provides us a rough estimate of the transition path region. Despite the fact they can give us a rough estimation of the transition path region, none of them can point the transition path region correctly. Moreover, it is very difficult to get the transition regions with these methods when the complexity of the system increases.

In the present work, we have tried Monte-Carlo simulation and Brownian dynamics data to locate the transition region of protein folding. Sampling the trajectories, which resulted in successful transitions, enabled us to pin point the start and end of the transition region correctly. The start and the end of the transition region converged with the increase in the number of trajectories sampled in the Monte-Carlo simulation and by increasing the number of data points in the Brownian dynamics simulation. One of the main problems in detecting the intermediate states in the folding trajectories is minimum fluctuations present in the system. This minimum fluctuation, which corresponds to ½ $k_BT$ energy, act as noise and hide the possible intermediate states present in the folding trajectories. This makes the direct observation of these intermediate states very difficult using experimental techniques. This is mainly because of the time resolution of these experimental techniques and the inherent noise present in the experimental setup. Previously, statistical methods have been applied on the single-molecule FRET trajectories to estimate the intermediate states present in the folding trajectories [22]. Our method gives us the increase in the transition region with respect to the minimum fluctuations present in the system. This information may help in detecting the possible intermediate states present in folding trajectories.

**Robustness of the proposed method**

We observed that with increase in the number of sampled trajectories (in case of Monte-Carlo simulations) and for longer time run i.e. with increased data points (in case of Brownian dynamics simulation), the transition path region converged to a fixed region in the energy profile along the reaction coordinate. In case of Brownian dynamics simulation for symmetric potential, the number of trajectories starting from the range [-2.1, -1.9] (for unfolding trajectories) and [1.9, 2.1] (for folding trajectories) and finally resulted in a successful transition increased monotonically with increase in the run time. But in case of asymmetric potentials, the trajectories which started from the above range and resulted in a successful transition varied with the run time (see S4, supplementary text). The cause of this variation is evident from the probability distribution of the protein in the asymmetric potentials. As the protein is spending more time diffusing in the well corresponding to the native state, a trajectory which is starting from this range may diffuse back to this range without contributing into a successful transition. Since, the diffusion in the well corresponds to the native state increases as the asymmetry in the potential increases, the probability of a trajectory starting from the above range and resulting in a successful transition decreases with the increasing asymmetry in the potential.

In case of Monte-Carlo simulation all trajectories were starting from the native (for unfolding transitions) and unfolded state (for folding transition). But in case of Brownian dynamic simulation, the trajectories which resulted in successful transitions were extracted from the Brownian dynamics data which was used for estimating the kinetic properties of the protein. Since the probability that a trajectory will exactly start from the native and unfolded state in the simulation data and will finally result into a successful transition is very small. So, we require a range from where these trajectories may start. The range that we have taken for our study is [-2.1, -1.9] for unfolding transitions and [1.9, 2.1] for folding transitions. We have found that the obtained transition region does not depend on the choice of this extension range for starting of these trajectories (see S5, supplementary text).

There are the two aspects through which our study can be viewed. It can be viewed as the study of different proteins in the same environmental conditions or the study of the same protein in different environmental conditions. Previous studies have shown that the environmental conditions play a very vital role in deciding the kinetics of the fibril formation [23-25]. In our study we have found that there is two order difference between the diffusion coefficient obtained from Kramers' theory and the average apparent diffusion coefficients obtained from the simulation itself. The percentage of time the protein spending in the transition region are 1.05%, 1.08% and 2.13% respectively which increased as the asymmetry of the potential increased (see Table. 1). That means there may be some hidden states in the transition path region. Due to those states the protein is spending a significant amount of time in the transition path region before making a successful transition where it is having some different conformations. These are the conformations that decide the folding behaviour and functional shape of the protein. Due to some change in the environmental condition if the protein gets misfolded that may cause aggregation. The fact that amyloids are formed in a particular environmental condition reflects the specificity of the environmental condition for a protein to act as a nucleation seed. Studying the behaviour of the proteins in both amyloid forming and non-amyloid forming environments will help us in understanding the difference in the behaviour of protein at the microscopic level in those conditions. Also, there are many proteins which don't form amyloids. One of the reasons is their ability to self chaperon against aggregation by limiting their conformational freedom [26]. In our study, by increasing the asymmetry we are changing the conformational freedom of the protein. Thus, this study can be used to get the insight of the dependence of non-amyloid forming proteins on their conformational freedom for self chaperon process.

## Conclusion

In this study, we have proposed a new method to estimate the transition path region which is a key parameter for a protein folding pathway. Monte-Carlo and Brownian dynamic simulation data were analysed by our new approach to locate the transition path region and their robustness were checked in different range. The method is based on the fact that the pre-transition and post-transition points should be distributed about the starting point of the transition region corresponding to both native and unfolded regions of the potential profile. This method enables us to get the transition region correctly which was not possible with the earlier two methods used for estimating the transition path region (see the main text). From the distribution of the points (see Fig. 2), it is clear that the points should be evenly distributed over a range with mean being the point corresponding to the beginning of the transition region in both native and unfolded part of the potential. We observed that the transition region expands with the increase in the asymmetry of the double well potential. The kinetic parameters, diffusion constant and transition path time were estimated by using Kramers' theory. The apparent diffusion in the native state increased with the increase in the asymmetry of the double well potential. This is mainly because of the increased conformational freedom of the protein in the native state. The diffusion coefficient obtained from Kramer's theory is two orders faster than the obtained apparent diffusion coefficients.

## Acknowledgement

DD and AS acknowledges the support of institute fellowship and ANG acknowledges the support of ISIRD grant from the Indian Institute of Technology Kharagpur, Kharagpur, India.

# Supplementary Information

## Locating transition path region in the pathway of protein folding

Debajyoti De, Anurag Singh and Amar Nath Gupta*

Biophysics and Soft Matter Laboratory, Department of Physics, IIT Kharagpur, Kharagpur-721302, India

*Corresponding author's email id: ang@phy.iitkgp.ernet.in

**S1. Convergence of transition path region**

We observed that with increase in the number of sampled trajectories (in case of Monte-Carlo (MC) simulation) and for longer simulation run time i.e. with increased data points (in case of Brownian dynamics (BD) simulation), the transition path region converged to a fixed region in the energy profile along the reaction coordinate. In case of MC simulation, even though the analysis was done for 10,000 trajectories, we obtained sufficient convergence for 1000 trajectories itself. Let *n* be the number of sampled trajectories in case of MC simulation and *m* be the simulation run time in case of BD simulation. We checked the variation of the transition path region with the increase in *n* and *m* values. Figure.1 shows the convergence of the transition path region obtained using MC simulation and BD simulation with increasing *n* and *m* values respectively.

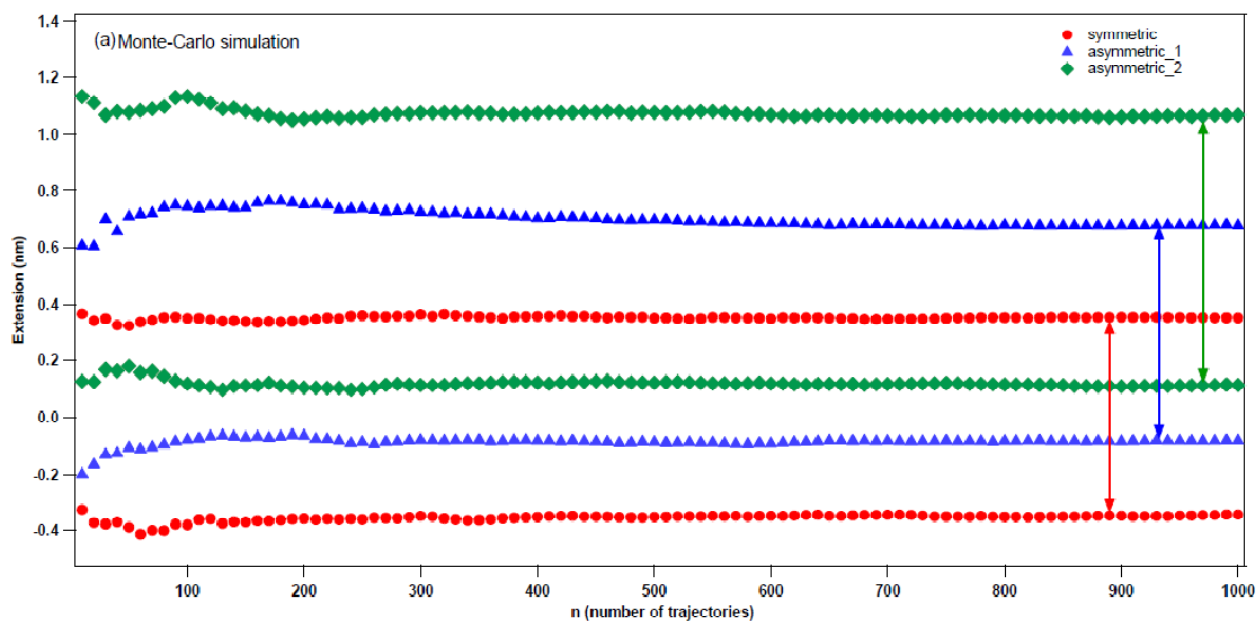

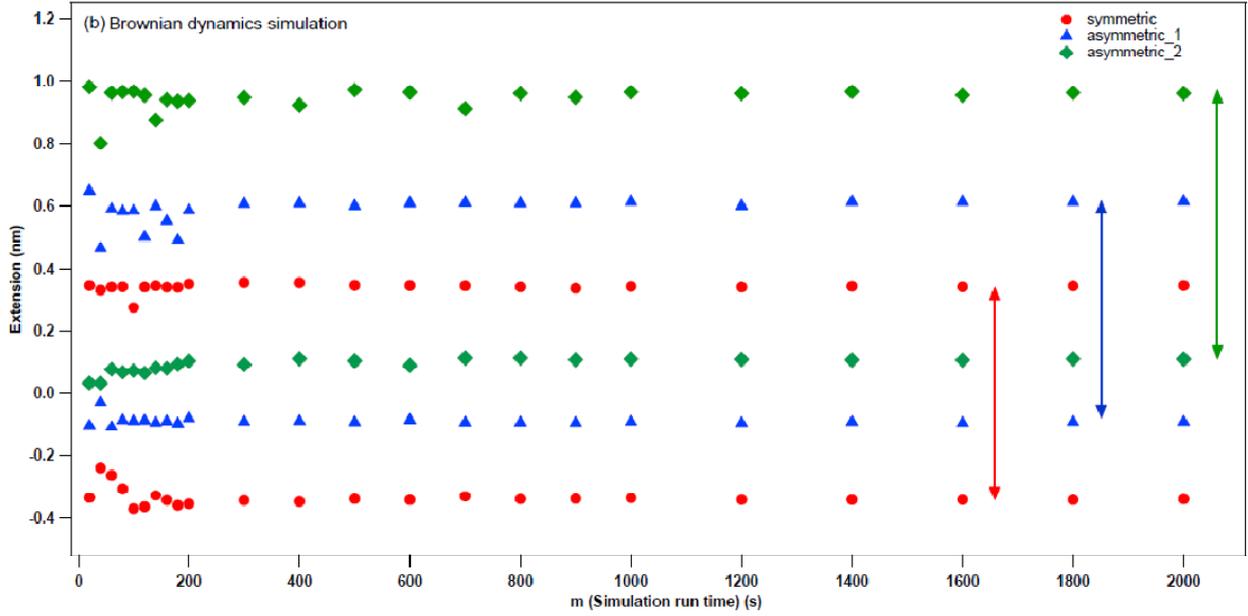

**Figure.1**: The red, blue and green colours correspond to the variation in transition path region for symmetric, asymmetric_1 and asymmetric_2 potentials respectively. The arrows follow the same colour code to show the transition path region for the three potentials. The distance between the two ends of the arrow gives the transition path region for the corresponding potential. (a) **MC simulation**: Convergence of transition path region with increase in the number of sampled trajectories. (b) **BD simulation**: Convergence of transition path region with increase in simulation run time i.e. with increase in the number of data points.

### S2. Distribution of pre-transition points about energy barrier

We observed that the pre-transition points followed specific distributions about the energy barrier peak. The nature of the distribution of these points depends on the slope of the potential profile. In case of symmetric potential, the pre-transition points followed a Gaussian distribution for both folding and unfolding transitions. This is due to the same slope of the potential on both the side of the energy barrier. With the increasing asymmetry in the potential, the distribution changes from Gaussian to exponentially growing for unfolding transitions and exponentially decaying for the folding transitions. Also, the rates of growth and decay of the exponential distribution depends on the slope of the potential. This is evident from the decay rates of the distribution tabulated in table1.

**Table 1**. The fitting parameters for the pre-transition distributions for symmetric and asymmetric_1 potentials obtained from MC and BD simulations.

| Result obtained from MC simulation | | | | Result obtained from BD simulation | | | |
|---|---|---|---|---|---|---|---|
| Symmetric | | Asymmetric_1 | | Symmetric | | Asymmetric_1 | |
| $\sigma$ (Left) | $\sigma$(Right) | $\tau^{-1}$(Left) | $\tau^{-1}$(Right) | $\sigma$ (Left) | $\sigma$(Right) | $\tau^{-1}$(Left) | $\tau^{-1}$(Right) |
| 0.55 | 0.55 | -1.57 | 3.44 | 0.61 | 0.60 | -1.55 | 3.96 |

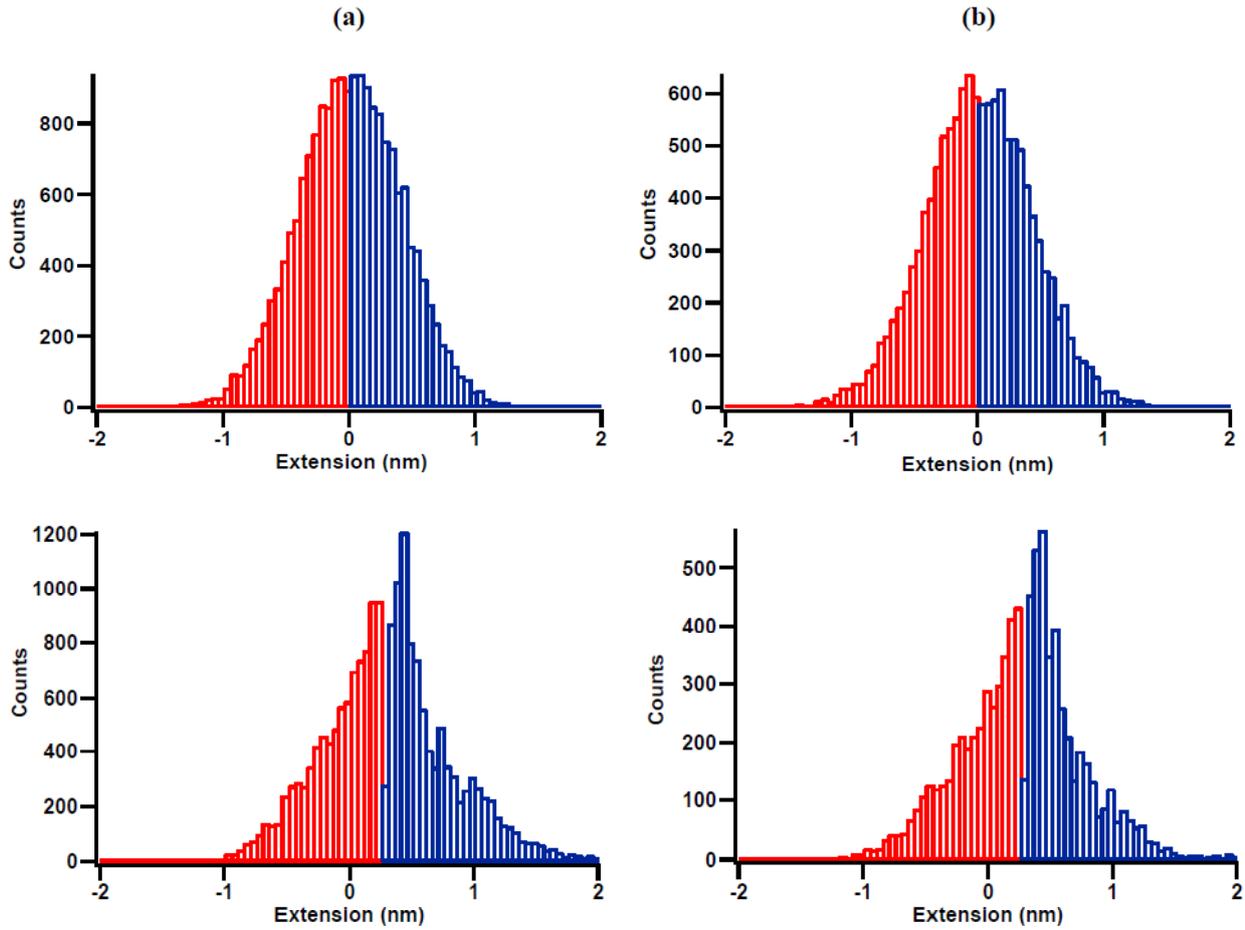

**Figure 2**: The panel (a) and (b) shows the distribution of pre-transition points in the native and unfolded state of the potential obtained using MC and BD simulations respectively for symmetric and asymmetric_1 potential. The red (blue) colour corresponds to the pre-transition points for unfolding (folding) transitions.

With the increase in the asymmetry of the double well potential, the slope corresponding to the native (unfolded) state becomes less (more) steep. This is evident from the growth and decay rate of the pre-transition points in the native and unfolded state for symmetric and asymmetric_1 potential (see table1). The Figure 3 shows the distribution of the pre-transition points for unfolding and folding transitions for symmetric and asymmetric_1 potential. For asymmetric_2 potential, distribution of similar nature was obtained but with a smaller (larger) growth (decay) rate of pre-transition points for unfolding (folding) transitions (not shown here).

**S3. Previously used method for determining the transition path region**

We have also used two of the methods that people have used in past for estimating the transition path region. Firstly we have fitted the barrier of symmetric and asymmetric double well potential with Gaussian and skew-Gaussian function to determine the transition path region. The symmetric potential is fitted with the function $y_0 + A\exp\left\{-\left(\frac{x-x_0}{width}\right)^2\right\}$. The values of the parameter we chose during the fitting are $y_0 = 1.69$, $A = 8.51$, $x_0 = 0$, width $= 1.29$. The asymmetric potential is fitted with function $y_0 + A\exp\left\{\frac{-(x-x_0)^2}{2(\sigma-c(x-x_0))^2}\right\}$. The values of the parameters for the asymmetric_1 and asymmetric_2 are tabulated below.

**Table 2.** The values of the fitting parameters of the skew-Gaussian function for both the asymmetric potentials

|  | $y_0$ | $A$ | $x_0$ | $\sigma$ | $c$ |
|---|---|---|---|---|---|
| Asymmetric_1 | 3.1461 | 6.10 | 0.17967 | -0.02176 | 1.9503 |
| Asymmetric_2 | 0.42 | 9.5 | 0.54 | 2.1 | -1.45 |

The points at which the curve started leaving the potential give us the transition path region. The TP region we obtained by fitting the double well potentials are [-0.45, 0.45], [0.06, 0.49], [0.09, 0.75] for symmetric and asymmetric respectively. But this point depends highly on the standard deviation of the Gaussian and skew-Gaussian curve. The standard deviation depends on the fluctuations present in the system which is very difficult to quantify accurately. As a result, it is very difficult to pin point the transition path region correctly using this fitting method.

Secondly the transition path region was identified observing the trajectories of the protein molecule between the folding and unfolding under a constant force. From the trajectory of the protein molecule between the folded and unfolded states we can predict the transition region (see Fig. 5).

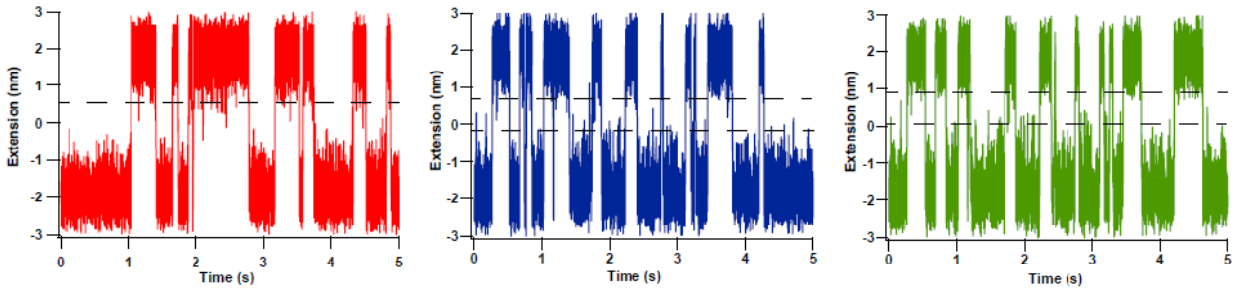

**Figure 3**: The fluctuation of the protein between the folded and unfolded state is shown. The dashed line shows TP regions for symmetric, asymmetric_1 and asymmetric_2 double well potentials respectively.

The regions we obtained from this rough estimation are [-0.48, 0.48], [-0.06, 0.68], [0.09, 0.75]. But the above method is very crude and from this method we can only have a rough estimation of the transition path region. But, none of them can correctly identify the transition path region.

## S4. Number of unfolding transitions as qualitative estimate of difference in transition probability

In the main text, we reported that with the increase in the asymmetry of the potential the probability of finding the protein in the native state increases. The analysis done for finding the transition path region with the proposed method further confirms the inference that this is due to the increased degree of freedom of the protein in the native state which allows it to diffuse more in the native state before making a successful transition. For this analysis, we extracted the number of unfolding transitions from BD simulation data with different simulation run time. Figure 4 shows the variation of the number of unfolding transitions with the increase in the simulation run time. From the figure it is clear that numbers of unfolding transitions are less for asymmetric potentials than the symmetric potentials.

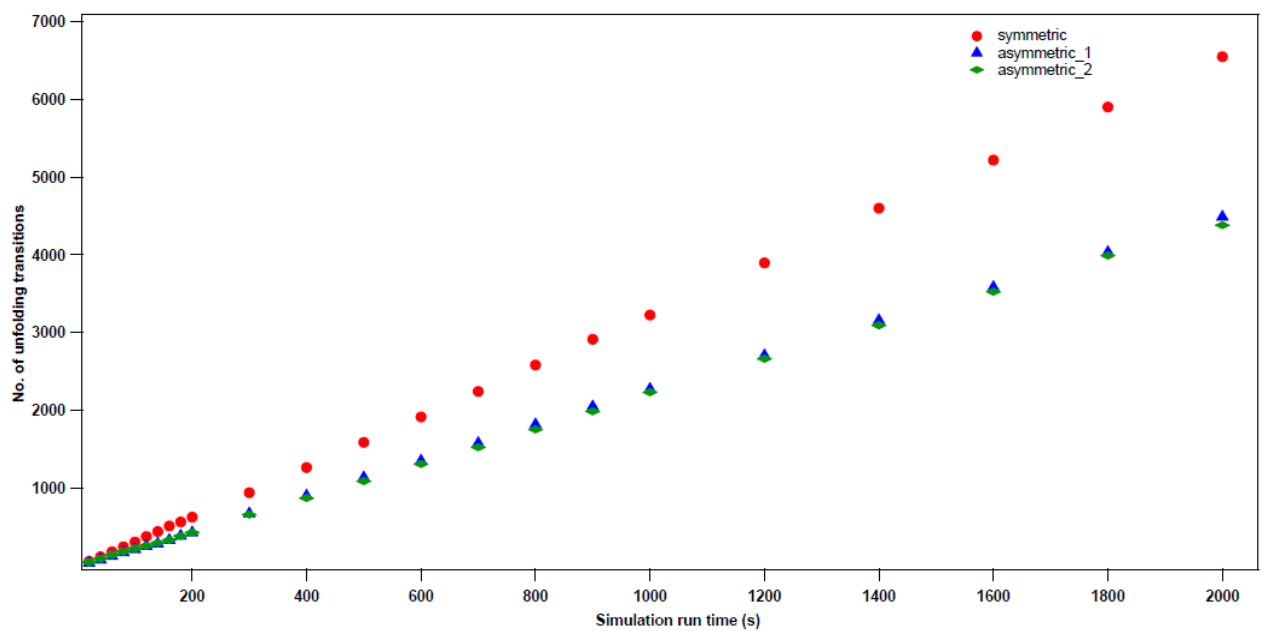

**Figure 4**: The variation of number of unfolding transitions with simulation run time for BD simulation. The red, blue and green colours correspond to the symmetric, asymmetric_1 and asymmetric_2 potentials respectively.

### S5. Dependence of transition path region on starting extension range for the trajectories in case of BD simulation

In case of MC simulation all trajectories were starting from the native (for unfolding transitions) and unfolded state (for folding transition). But in case of BD simulation, the trajectories which resulted in successful transitions were extracted from the BD data which was used for estimating the kinetic properties of the protein. Since, the probability that a trajectory will exactly start from the native and unfolded state in the simulation data and will finally result into a successful transition is very small. Hence, we required a range from where these trajectories may start. The range that we had taken for our study was [-2.1, -1.9] for unfolding transitions and [1.9, 2.1] for folding transitions. Fig. 2 shows that the obtained transition region does not depend on the choice of this extension range for starting of these trajectories.

In Fig.2, the extension ranges from where the trajectories are starting are shown with a range index. The extension ranges corresponding to theses range indices are: 0 – [-2.01, -1.99] & [1.99, 2.01], 1 – [-2.02, -1.98] & [1.98, 2.02], 2 – [-2.03, -1.97] & [1.97, 2.03], 3 – [-2.04, -1.96] & [1.96, 2.04], 4 – [-2.05, -1.95] & [1.95, 2.05], 5 – [-2.06, -1.94] & [1.94, 2.06], 6 – [-2.07, -1.93] & [1.93, 2.07], 7 – [-2.08, -1.92] & [1.92, 2.08], 8 – [-2.09, -1.91] & [1.91, 2.09], 9 – [-2.1, -1.9] & [1.9, 2.1], 10 – [-2.11, -1.89] & [1.89, 2.11], 11 – [-2.12, -1.88] & [1.88, 2.12], 12 – [-2.13, -1.87] & [1.87, 2.13], 13 – [-2.14, -1.86] & [1.86, 2.14], 14 – [-2.15, -1.85] & [1.85, 2.15] respectively. To each range index there are two extension ranges corresponding to the unfolding and folding transitions respectively.

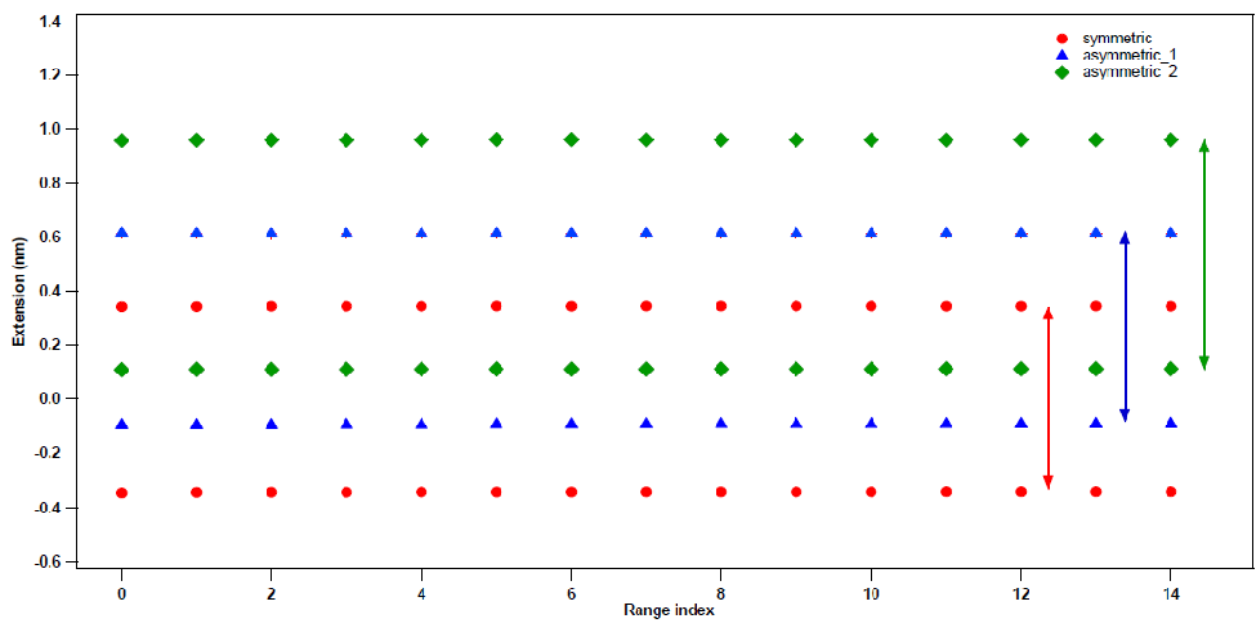

**Figure.5**: The red, blue and green colours correspond to the variation in transition path region for symmetric, asymmetric_1 and asymmetric_2 potentials respectively. The arrows follow the same colour code to show the transition path region for the three potentials. Each range index corresponds to a range along the reaction coordinate from where the trajectories are starting. The distance between the two ends of an arrow gives the transition path region for the corresponding potential. The range corresponding to the range indices are: 0 – [-2.01, -1.99] & [1.99, 2.01], 1 – [-2.02, -1.98] & [1.98, 2.02], 2 – [-2.03, -1.97] & [1.97, 2.03], 3 – [-2.04, -1.96] &[1.96, 2.04], 4 – [-2.05, -1.95] &[1.95, 2.05], 5 – [-2.06, -1.94] &[1.94, 2.06], 6 – [-2.07, -1.93] &[1.93, 2.07], 7 – [-2.08, -1.92] &[1.92, 2.08], 8 – [-2.09, -1.91] &[1.91, 2.09], 9 – [-2.1, -1.9] &[1.9, 2.1], 10 – [-2.11, -1.89] &[1.89, 2.11], 11 – [-2.12, -1.88] &[1.88, 2.12], 12 – [-2.13, -1.87] &[1.87, 2.13], 13 – [-2.14, -1.86] &[1.86, 2.14], 14 – [-2.15, -1.85] &[1.85, 2.15] respectively.

S1, S2, S4 and S5 indicate towards the robustness of our method. We compared the results obtained in S3 with our results. S3 shows the reliability of our method in comparison to the previously used methods for the estimation of transition path region.